\documentstyle[aps,prl,draft]{revtex} 
\begin{document}
\draft

\def\annil{$A+A\to0$}
\def\coal{$A+A\to A$} 
\def \erf{{\rm erf}}
\def\erfc{{\rm erfc}}
 
\title{Correlation functions for diffusion-limited annihilation, \annil}
\author{Thomas O. Masser and Daniel ben-Avraham\footnote
  {{\bf e-mail:} benavraham@clarkson.edu}}

\address{Physics Department, and Clarkson Institute for Statistical
Physics (CISP), \\ Clarkson University, Potsdam, NY 13699-5820}
\maketitle

\begin{abstract} 
The full hierarchy of multiple-point correlation functions for
diffusion-limited annihilation, \annil, is obtained
analytically and explicitly, following the method of intervals.  In the long
time asymptotic limit, the correlation functions of annihilation are identical
to those of coalescence, \coal, despite differences between the two models
in other statistical measures, such as the interparticle
distribution function.
\end{abstract}

The kinetics of nonequilibrium processes, in particular
diffusion-limited reactions, have attracted much recent
interest~\cite{Benson60,Laidler65,Haken78,Nicolis80,vanKampen81,Ligget85,%
reaction-reviews,JStatPhys}.  
Because of the lack of a comprehensive approach for the study of such systems,
models that yield to an exact analysis are of prime importance.  In this
respect, none have been studied more than diffusion-limited annihilation,
\annil~\cite{Bramson80,Torney83,Racz85,Lushnikov87,Kopelman88,Spouge88,%
Balding89,%
Amar90,Krapivsky93,Alcaraz94,Grynberg94,Alemany95,Henkel95,Schutz95,Krebs95,%
Bares99,Park00,Masser00,Lindenberg01,Khorrami01}, and
coalescence,
\coal~\cite{Bramson80,Kopelman88,Spouge88,Balding89,Krapivsky93,%
Henkel95,Krebs95,Masser00,Khorrami01,Takayasu88,DbA90,Doering91,%
DbA98,Privman94}. 
Known exact results include the time dependence of the
particle concentration and the two-point correlation function (for finding two
particles at two different points, simultaneously).  It has also been shown
that the full hierarchy of $n$-point correlation functions for
the two processes is identical~\cite{Henkel95,Krebs95,Balboni95,Simon95}, but
explicit expressions for $n>3$ are unavailable. 

Here we attack the problem of
correlation functions for annihilation, using the method
of parity intervals (or even/odd
intervals)~\cite{Alemany95,Masser00,Lindenberg01,DbA00}.  We recover the
identity relation of the $n$-point correlation functions for annihilation and
coalescence, and we derive {\em explicit} expressions, valid in the long time
asymptotic limit, for all
$n$.  

Consider the annihilation model, 
defined on the line
$-\infty<x<\infty$.  Particles $A$ are represented by points which perform
unbiased diffusion with a diffusion constant $D$. When two particles meet
they annihilate instantly.  Since the reaction step is infinitely
fast, the system models the {\it diffusion-limited\/} annihilation process
\annil.

An exact treatment of the problem is possible through the method of parity
intervals~\cite{Alemany95,Masser00,Lindenberg01,DbA00}.  The key parameter
is
$G(x,y;t)$---the probability that the interval
$[x,y]$ contains an even number of particles at time~$t$~\cite{remark}. 
Particles near the  edges of an interval may diffuse into or out of the
interval, affecting the probability
$G$.  (On the other hand, reactions inside the interval cannot affect its
parity.)  With this observation in mind, one derives a rate equation for
the probability $G(x,y;t)$~\cite{Masser00}:
\begin{equation}
\label{dG/dt}
\frac{\partial}{\partial t} G(x,y;t)=D(\frac{\partial^2}{\partial x^2}
  +\frac{\partial^2}{\partial y^2})G(x,y;t)\;.
\end{equation}
The annihilation reaction imposes the boundary condition
\begin{mathletters}
\label{bc:G}
\begin{equation}
\label{bc:G=1}
\lim_{x\uparrow y{\rm\ or\ }y\downarrow x}G(x,y;t)=1\;,
\end{equation}
and $G$ must also obey the conditions required from a probability
density function.  If the initial distribution of particles is random, then
$G(x,y;0)=\case{1}{2}+\case{1}{2}\exp[-2c_0(y-x)]$, where $c_0$ is their
initial density.  In this case we have the additional boundary condition
\begin{equation}
\label{bc:Ginfty}
\lim_{x\to-\infty{\rm\ or\ }y\to\infty} G(x,y;t)=\case{1}{2}\;.
\end{equation}
\end{mathletters}
From $G(x,y;t)$ one can derive the particle concentration:
\begin{equation}
\label{rho1}
\rho(x;t)=-{\partial\over\partial y}G(x,y;t)|_{y=x}\;.
\end{equation}

Let $\rho_n(x_1,x_2,\dots,x_n;t)$ be the $n$-point density correlation
function, for finding particles at each of the locations
$x_1,x_2,\cdots,x_n$ at time
$t$.  The particle concentration, $\rho(x,t)\equiv\rho_1(x,t)$, represents
merely the first term in the hierarchy $\{\rho_n\}$, ${n=1,2,\dots}$.  

The correlation functions may be obtained from a
generalization of the method of parity intervals, in the following way.  Let
$H_n(x_1,y_1,\overline{x_2,y_2},\dots,x_n,y_n;t)$ be the joint probability that
the interval $[x_1,y_1]$ contains an even number of particles, $[x_2,y_2]$
contains an odd number, etc., (odd intervals are denoted by an overbar), at
time~$t$.  The intervals are non-overlapping, and ordered:
$x_1<y_1<\cdots<x_n<y_n$.  Let $F_n(x_1,y_1,x_2,y_2,\dots,x_n,y_n;t)$
denote the probability that the total number of particles contained in
$\bigcup_{i=1}^n[x_i,y_i]$, is even.  Thus,
\begin{mathletters}
\begin{eqnarray}
F_1(x_1,y_1;t)&=&H_1(x_1,y_1;t)=G(x_1,y_1;t)\;,\\
F_2(x_1,y_1,x_2,y_2;t)&=&H_2(x_1,y_1,x_2,y_2;t)
  + H_2(\overline{x_1,y_1},\overline{x_2,y_2};t)\;,
\end{eqnarray}
\end{mathletters}
and, in general, $F_n$ is expressible as a sum of $2^{n-1}$\ \ $H_n$ functions,
corresponding to the different combinations of interval parities that
contribute to a total number of particles that is even. Then, in view of
Eq.~(\ref{rho1}), the
$n$-point correlation function is given by
\begin{equation}
\label{rho_n}
\rho_n^{\rm anni}(x_1,\dots,x_n;t)=
\frac{(-1)^n}{2^{n-1}}\frac{\partial^n}{\partial y_1\cdots\partial y_n}
F_n(x_1,y_1,\dots,x_n,y_n;t)|_{y_1=x_1,\dots,y_n=x_n}\;.
\end{equation}
The $H_n$ satisfy a 
$2n$-dimensional diffusion equation, analogous to Eq.~(\ref{dG/dt}), and for
similar reasons.  However, the boundary conditions of this equation are
complicated by the following fact. For
$y_i\to x_{i+1}$, a particle moving from
the $i$-th interval to the $(i+1)$-th interval, or vice versa, would flip
the parity of the two adjacent intervals.  On the other hand, $F_n$ satisfies
the same diffusion equation as $H_n$,
\begin{equation}
\label{dFn/dt}
\frac{\partial}{\partial t} F_n(x_1,y_1,\dots,x_n,y_n;t)
=D(\frac{\partial^2}{\partial x_1^2}+\frac{\partial^2}{\partial y_1^2} + \cdots
+\frac{\partial^2}{\partial x_n^2} + \frac{\partial^2}{\partial y_n^2})F_n\;,
\end{equation}
but the boundary conditions are simpler: $F_n$ contains also the case
where the parity of the intervals $i$ and $(i+1)$ is flipped, so it is not
affected by a particle hopping between the two intervals.  If
interval $i$ is shrunk to zero, 
Eq.~(\ref{bc:G=1}) yields the boundary condition
\begin{mathletters}
\label{bc:Fn}
\begin{equation}
\lim_{x_i\uparrow y_i{\rm\ or\ }y_i\downarrow x_i}
  F_n(x_1,y_1,\dots,x_n,y_n;t)=
  F_{n-1}(x_1,y_1,\dots,\not{\!x_i},\not{\!y_i},\dots,x_n,y_n;t)\;,
\end{equation}
where we use the notation that crossed out arguments (e.g.
${\not{\!x_i}}$) have been removed.  
If the endpoints of two adjacent intervals are brought together, the intervals
merge, resulting in the boundary condition
\begin{equation}
\lim_{y_i\uparrow x_{i+1}{\rm\ or\ }x_{i+1}\downarrow y_i}
 F_n(x_1,y_1,\dots,x_n,y_n;t)=
 F_{n-1}(x_1,y_1,\dots,\not{\!y_i},\not{\!x_{i+1}},\dots,x_n,y_n;t)\;.
\end{equation}
Finally, for a random initial
distribution of particles, we have
\begin{equation}
\lim_{x_1\to-\infty{\rm\ or\ }y_n\to\infty}F_n(x_1,y_1,\dots,x_n,y_n;t)=
  \case{1}{2}\;.
\end{equation}
\end{mathletters} 
The
$F_n$ are tied together in an hierarchical fashion through the boundary
conditions~(\ref{bc:Fn}a) and (\ref{bc:Fn}b):  one must know $F_{n-1}$ in
order to compute $F_n$.  At the root of the hierarchy, $F_1=G$ is obtained from
Eqs.~(\ref{dG/dt}) and (\ref{bc:G}).

The problem posed by Eqs.~(\ref{dG/dt}), (\ref{bc:G}), (\ref{dFn/dt}),
(\ref{bc:Fn}) is similar to that of diffusion-limited coalescence,
\coal~\cite{DbA98}.  In that case one defines $E_n(x_1,y_1,\dots,x_n,y_n;t)$ as
the joint probability of finding the intervals $[x_i,y_i]$, $i=1,2,\dots,n$,
{\em empty} at time $t$.  $E_1(x,y;t)\equiv E(x,y;t)$ satisfies the same
equation as Eq.~(\ref{dG/dt}):
\begin{equation}
\label{dE/dt}
\frac{\partial}{\partial t} E(x,y;t)=D(\frac{\partial^2}{\partial x^2}
  +\frac{\partial^2}{\partial y^2})E(x,y;t)\;,
\end{equation}
with the boundary conditions
\begin{mathletters}
\label{bc:E}
\begin{eqnarray}
\label{bc:E=1}
\lim_{x\uparrow y{\rm\ or\ }y\downarrow x}E(x,y;t)&=&1\;,\\
\label{bc:Einfty}
\lim_{x\to-\infty{\rm\ or\ }y\to\infty}E(x,y;t)&=&0\;.
\end{eqnarray}
\end{mathletters}
Note the difference between the boundary conditions (\ref{bc:Ginfty}) and
(\ref{bc:Einfty}). Likewise, $E_n$ satisfies the same equation as
Eq.~(\ref{dFn/dt}): 
\begin{equation}
\label{dEn/dt}
\frac{\partial}{\partial t} E_n(x_1,y_1,\dots,x_n,y_n;t)
=D(\frac{\partial^2}{\partial x_1^2}+\frac{\partial^2}{\partial y_1^2} + \cdots
+\frac{\partial^2}{\partial x_n^2} + \frac{\partial^2}{\partial y_n^2})E_n\;,
\end{equation}
with boundary conditions analogous to~(\ref{bc:Fn}a) and
(\ref{bc:Fn}b), 
\begin{mathletters}
\label{bc:En}
\begin{eqnarray}
\lim_{x_i\uparrow y_i{\rm\ or\ }y_i\downarrow x_i}
  E_n(x_1,y_1,\dots,x_n,y_n;t)&&=
  E_{n-1}(x_1,y_1,\dots,\not{\!x_i},\not{\!y_i},\dots,x_n,y_n;t)\;,\\
\lim_{y_i\uparrow x_{i+1}{\rm\ or\ }x_{i+1}\downarrow y_i}
 E_n(x_1,y_1,\dots,x_n,y_n;t)&&=
 E_{n-1}(x_1,y_1,\dots,\not{\!y_i},\not{\!x_{i+1}},\dots,x_n,y_n;t)\;.
\end{eqnarray}
but  
\begin{equation}
\label{bc:Eninfty}
\lim_{x_1\to-\infty{\rm\ or\ }y_n\to\infty}E_n(x_1,y_1,\dots,x_n,y_n;t)=
  0\;,
\end{equation}
\end{mathletters}
instead of~(\ref{bc:Fn}c).  The $n$-point correlation function for coalescence
is
\begin{equation}
\label{rho_n.coal}
\rho_n^{\rm coal}(x_1,\dots,x_n;t)=
(-1)^n\frac{\partial^n}{\partial y_1\cdots\partial y_n}
E_n(x_1,y_1,\dots,x_n,y_n;t)|_{y_1=x_1,\dots,y_n=x_n}\;.
\end{equation}

Eqs.~(\ref{dG/dt}), (\ref{bc:G}), (\ref{dFn/dt}), (\ref{bc:Fn}),
and~(\ref{dE/dt}) -- (\ref{bc:En}) imply that the
solutions for $F_n$ and
$E_n$ are simply related:
\begin{equation}
\label{FnEn}
F_n(x_1,y_1,\dots,x_n,y_n;t)=\case{1}{2}
   +\case{1}{2}E_n(x_1,y_1,\dots,x_n,y_n;t)\;,
\end{equation}
provided that the same relation holds also for the initial conditions. 
Suppose that the initial distribution of particles is random, with initial
concentration $c_0^{\rm anni}$ for annihilation, and $c_0^{\rm coal}$ for
coalescence.  Then
$E_n(x_1,\dots,y_n;0)=\exp[-c_0^{\rm coal}(y_1-x_1+\cdots+y_n-x_n)]$, while
$F_n(x_1,\dots,y_n;0)=\case{1}{2}+\case{1}{2} 
\exp[-2c_0^{\rm anni}(y_1-x_1+\cdots+y_n-x_n)]$.  Thus, the
relation~(\ref{FnEn}) is satisfied if $c_0^{\rm
anni}=\case{1}{2}c_0^{\rm coal}$.  Moreover, Eqs.~(\ref{rho_n}) and
(\ref{rho_n.coal}) imply that in this case
\begin{equation}
\label{equiv}
\rho_n^{\rm anni}(x_1,\dots,x_n;t)=
  (\case{1}{2})^n\rho_n^{\rm coal}(x_1,\dots,x_n;t)\;,
\end{equation}
for all $n$.  In other words, {\em the {\rm n}-point correlation
functions for annihilation and coalescence are identical}, as already found by
others~\cite{Henkel95,Krebs95,Balboni95,Simon95}.  

We now produce explicit expressions for the $n$-point correlation functions in
the long time asymptotic limit.  Recall first the solution for
$E_n$. For $n=2$ the
solution is~\cite{unpublished},
\begin{eqnarray}
\label{F2}
&&E_2(x_1,y_1,x_2,y_2;t)= \nonumber\\
&&E(x_1,y_1;t)E(x_2,y_2;t)-E(x_1,x_2;t)E(y_1,y_2;t)
  +E(x_1,y_2;t)E(y_1,x_2;t)\;,
\end{eqnarray}
where $E(x,y;t)$ is the solution of Eqs.~(\ref{dE/dt}), (\ref{bc:E}).
Generally, for~$n\geq2$~\cite{DbA98},
\begin{equation}
\label{En}
E_n(x_1,y_1,\dots,x_n,y_n;t) = 
  \sum_{p=1}^{(2n-1)!!}\sigma_pE(z_{1,p},z_{2,p};t)E(z_{3,p},z_{4,p};t)\cdots 
  E(z_{2n-1,p},z_{2n,p};t)\;,
\end{equation}
where $z_{1,p},z_{2,p},\dots,z_{2n,p}$ is an {\it ordered\/}
permutation, $p$, of the variables $x_1,y_1,\dots,x_n,y_n$, such that
\begin{equation}
\label{order}
z_{1,p}<z_{2,p},\;z_{3,p}<z_{4,p},\dots,z_{2n-1,p}<z_{2n,p},\quad{\rm and}\quad 
z_{1,p}<z_{3,p}<z_{5,p}\cdots<z_{2n-1,p}\;.
\end{equation}
There are exactly $(2n-1)!!=1\cdot3\cdot\,\cdots\,\cdot(2n-1)$
such permutations. $\sigma_p$ is $+1$ for even permutations
(permutations that require an even number of exchanges between pairs of
variables), or $-1$ for odd permutations.
Alternatively, the $E_n$ may be obtained through the recursion relation:
\begin{eqnarray}
\label{En.En-1}
E_n(x_1,y_1,\dots,x_n,y_n;t) =&& 
+\sum_{j=1}^nE(x_1,y_j;t)E_{n-1}(\not{\!x_1},y_1,\dots,x_j,\not{\!y_j},\dots,
  x_n,y_n;t) \nonumber\\
&&-\sum_{j=2}^nE(x_1,x_j;t)E_{n-1}(\not{\!x_1},y_1,\dots,\not{\!x_j},y_j,\dots,
  x_n,y_n;t) \;,
\end{eqnarray}
then $\rho_n$ may be computed through the relation~(\ref{rho_n.coal}),
or~(\ref{rho_n}) and (\ref{FnEn}).
  
Consider the long-time
asymptotic limit, where     
\begin{equation}
\label{Easymp}
E(x,y;t)=\erfc\big({y-x\over\sqrt{8Dt}}\big)\;.
\end{equation}
Then, the long-time asymptotic
$n$-point correlation function is:
\begin{mathletters}
\label{rho_n.asym}
\begin{equation}
\rho_n(x_1,y_1,\dots,x_n,y_n;t) 
  \mathop\to_{t\to\infty} (-\rho)^n
  \sum_{p=1}^{(2n-1)!!}\sigma_pC(z_{1,p},z_{2,p};t)C(z_{3,p},z_{4,p};t)\cdots 
  C(z_{2n-1,p},z_{2n,p};t)\;,
\end{equation}
where
\begin{equation}
\rho=\rho_1(x;t)=\cases{
{1/\sqrt{8\pi Dt}} & annihilation,\cr
{1/\sqrt{2\pi Dt}} & coalescence,
}
\end{equation}
\begin{equation}
C(z_1,z_2;t)=\cases{
-1 & $(z_1,z_2)=(x_k,y_k)\;,$\cr
\erfc(\xi_{lk}) & $(z_1,z_2)=(x_k,x_l)\;,$\cr
-e^{-\xi_{lk}^2} & $(z_1,z_2)=(x_k,y_l)\;,$\cr
e^{-\xi_{lk}^2} & $(z_1,z_2)=(y_k,x_l)\;,$\cr
-\sqrt{\pi}\xi_{lk}e^{-\xi_{lk}^2} & $(z_1,z_2)=(y_k,y_l)\;,$
}
\end{equation}
\end{mathletters}
and we used the notation $\xi_{lk}=(x_l-x_k)/\sqrt{8Dt}$.
For example, for $n=2,3$, we get the long-time asymptotic expressions:
\begin{mathletters}

\begin{equation}
\label{rho2}
{\rho_2(x_1,x_2;t)\over\rho^2}=
1-e^{-2\xi_{21}^2}+\sqrt{\pi}\,\xi_{21} e^{-\xi_{21}^2}\erfc(\xi_{21})\;,
\end{equation}
\begin{eqnarray}
\label{rho3}
{\rho_3(x_1,x_2,x_3;t)\over\rho^3} &=&
1-e^{-2\xi_{21}^2}-e^{-2\xi_{32}^2}-e^{-2\xi_{31}^2}
+2e^{-\xi_{21}^2-\xi_{32}^2-\xi_{31}^2}\nonumber\\
&&+\sqrt{\pi}\,\xi_{21} 
     (e^{-\xi_{21}^2}-e^{-\xi_{32}^2-\xi_{31}^2})
                               \erfc(\xi_{21})\nonumber\\ 
&&+\sqrt{\pi}\,\xi_{32}
     (e^{-\xi_{32}^2}-e^{-\xi_{21}^2-\xi_{31}^2})
                               \erfc(\xi_{32})\nonumber\\  
&&+\sqrt{\pi}\,\xi_{31}
     (e^{-\xi_{31}^2}-e^{-\xi_{21}^2-\xi_{32}^2})\erfc(\xi_{31})\;.
\end{eqnarray}
\end{mathletters}

In summary, we have confirmed the fact that the infinite hierarchies of
$n$-point correlation functions for coalescence and annihilation are
identical, using the method of parity intervals.  The simplicity of our
approach allowed us to obtain explicit expressions for the long-time asymptotic
limit, given in Eqs.~(\ref{rho_n.asym}). We note that our results are not
restricted to long times.  Indeed, for the case of a random distribution of
particles, such that the initial concentration for annihilation is half that
of coalescence, the identity holds at all times.  In this case, explicit
expressions for the correlation functions (valid at all times) can be obtained
by using the full solution of Eqs.~(\ref{dE/dt}), (\ref{bc:E}), 
\begin{equation}
E(x,y;t)=\erfc(\frac{y-x}{\sqrt{8Dt}})-\frac{1}{2}e^{2Dc_0^2t}
  \{e^{c_0(y-x)}[1-\erf(\frac{y-x+4Dc_0t}{\sqrt{8Dt}})]
  -e^{-c_0(y-x)}[1+\erf(\frac{y-x-4Dc_0t}{\sqrt{8Dt}})]\}\;,
\end{equation} 
instead of the asymptotic expression of
Eq.~(\ref{Easymp}).

Remarkably, the particle distributions in coalescence and annihilation
differ, despite the correspondence of the $n$-point correlation functions. 
The probability density function, $p(x)$, for the distance $x$ between two
neighboring particles illustrates this difference.  For large
$x$, $p(x)\sim e^{-x^2}$ for coalescence, while $p(x)\sim e^{-x}$ for
annihilation.  Evidently, the complete hierarchy of $n$-point correlation 
functions is not sufficient to determine an infinite-particle system
uniquely, and $p(x)$ cannot be computed from a knowledge of the
$\rho_n$.  A study of
$p(x)$ and $\{\rho_n\}$ in finite systems might illuminate this curious
phenomenon.

\acknowledgments
Funding of this project by the NSF (PHY-9820569) is
gratefully acknowledged.


\end{document}